# A Bi-channel Aided Stitching of Atomic Force Microscopy Images

Huanhuan Zhao[1,a], Ruben Millan-Solsona[2], Marti Checa[2], Spenser R. Brown[3], Jennifer L. Morrell-Falvey[3], Liam Collins[2], Arpan Biswas[4,b]

[1]Bredesen Center for Interdisciplinary Research, University of Tennessee, Knoxville, USA, 37996
[2]Center for Nanophase Materials Sciences, Oak Ridge National Laboratory, Oak Ridge, USA, 37830
[3]Biosciences Division, Oak Ridge National Laboratory, Oak Ridge, USA, 37830
[4]University of Tennessee-Oak Ridge Innovation Institute, University of Tennessee, Knoxville, USA, 37996

*Microscopy is an essential tool in scientific research, enabling the visualization of structures at micro- and nanoscale resolutions. However, the field of microscopy often encounters limitations in field-of-view (FOV), restricting the amount of sample that can be imaged in a single capture. To overcome this limitation, image stitching techniques have been developed to seamlessly merge multiple overlapping images into a single, high-resolution composite. The images collected from microscope need to be optimally stitched before accurate physical information can be extracted from post analysis. However, the existing stitching tools either struggle to stitch images together when the microscopy images are feature sparse or cannot address all the transformations of images when performing image stitching. To address these issues, we propose a bi-channel aided feature-based image stitching method and demonstrate its use on Atomic Force Microscopy (AFM) generated biofilm images as experimental data. The topographical channel image of AFM data captures the morphological details of the sample, and a stitched topographical image is desired for researchers. We utilize the amplitude channel of AFM data to maximize the matching features and to estimate the position of the original topographical images and show that the proposed bi-channel aided stitching method outperforms the traditional direct stitching approach in AFM topographical image stitching task. Furthermore, we found that the differentiation of the topographical images along the x-axis provides similar feature information to the amplitude channel image, which generalizes our approach when the amplitude images are not available. Here we demonstrated the application on AFM, but similar approaches could be employed of optical microscopy with brightfield and fluorescence channels. We believe this proposed workflow can serve as a valuable augmentation strategy for microscopy image stitching tasks and will benefit the experimentalist to avoid erroneous analysis and discovery due to incorrect stitching.*

[a] hzhao31@vols.utk.edu

[b] abiswas5@utk.edu

**Introduction**:

High-resolution microscopy is a cornerstone of scientific discovery, providing unparalleled insight into the structure and function of materials and biological specimens. However, a fundamental challenge in microscopy is the trade-off between resolution and field of view (FOV). High-magnification imaging reveals fine structural details but captures only a limited portion of the sample at a time. This constraint is particularly problematic in fields requiring both fine-scale resolution and large-area context, such as histopathology, neuroscience, and materials science. For example, in medical diagnostics[1–4], pathologists analyzing tissue biopsies must examine entire sections while still detecting subcellular abnormalities indicative of diseases like cancer. Similarly, in neuroscience[5–7], researchers studying brain tissue and spinal cord must visualize intricate neuronal networks spanning large regions while preserving the details of individual synaptic connections. Beyond biology, disciplines like astronomy[8,9] also face analogous challenges when stitching together high-resolution telescope images to create expansive sky maps. Similarly, to analyze geospatial data for predicting traffic conditions and optimizing transportation routes, large high-resolution satellite images need to be stitched together[10–13]. Here, seamless stitching of these high-resolution images is a crucial step to ensure 1) preservation of the physical insights of the original images and 2) avoiding potential false interpretation and discovery of the underlying information of the images. During the collection of these images, a degree of overlap between two adjacent images is set. If the degree is lower, the number of sampling images is reduced, which decreases the total scanning cost. While increasing the overlap, the data acquisition cost proportionally increases.

Different methods have been developed for effective image stitching, each with its own strengths and limitations. Direct methods, which use pixel intensities to align images, include techniques like Lucas-Kanade optical flow and the Sum of Absolute Differences (SAD). These methods are accurate for small displacements but can be computationally demanding and sensitive to illumination changes for large images. Transform-based methods, such as those utilizing the Fourier transform[14,15], determine translation between images through phase correlation, offering speed and noise resistance but these methods struggle with rotation and scaling. Multi-scale approaches, like pyramid-based stitching, create multi-resolution representations to improve robustness and accuracy through progressive alignment refinement. Feature-based methods[12,16,17] consider different feature detectors and feature matching algorithms in order to calculate the image transformation matrix. Scale-Invariant Feature Transform (SIFT)[18,19] and Speeded-Up Robust Features (SURF)[20] are two popular feature detectors. SIFT is known for its robustness to scale, rotation, and illumination changes, while SURF is optimized for speed but may be less effective under extreme variations. Other notable feature-detection algorithms are Binary Robust Invariant Scalable Keypoints (BRISK)[21] and Accelerated-KAZE (AKAZE)[22]. Oriented FAST and Rotated BRIEF (ORB)[23] offers computational efficiency and is suitable for real-time applications, though it handles scale variations less effectively than SIFT. Natural images, like buildings, mountains, or indoor scenes, typically contain rich texture details. These images are easily captured with cameras, and acquiring images with large overlaps is cost-effective, which simplifies the stitching task. Existing stitching tools, such as *AutoStitch*[16], *stitch2D*[24] and *Stitching*[25] are highly effective for stitching these types of images. In other words, these stitching methods generally work with a high volume of matching features obtained from significant amounts of image overlapping. However, microscopy images often have sparse features due to the nature of the sample and collecting highly overlapped images (say 30%) is not often feasible due to the cost and time-intensive experimental image acquisition process. In situations when image overlapping is limited, the performance of these stitching methods degrades significantly to the point that they are not even considered for any sort of post image analysis. Thus, as the stitching error increases between two adjacent images due to limited matching features (limited overlap), the error propagates and increases even more when trying to properly stitch non-adjacent images, making the stitching problem even harder to solve. Thus, there is a need for an improved image stitching method that can accurately stitch complex and noisy microscopy images with limited overlap due to experimental constraints.

Over the years, various image stitching techniques have been developed to computationally merge multiple microscopy images into seamless, high-resolution composites. When applying image stitching techniques to microscopy images, researchers have primarily adopted two approaches: feature-based methods[26–30] and Fourier-based methods[14]. Feature-based methods include feature detection, feature matching, camera pose estimation, warping and blending etc. In contrast, Fourier-based methods calculate the translational offsets between images using phase-correlation and then align the images based on these offsets. While Fourier-based techniques are straightforward and fast, they only address translational shifts, whereas feature-based approaches can handle additional transformations such as rotations and scaling. Ma et al. were among the first to apply a feature-based photography stitching software, *AutoStitch*, to microscopy images[31]. Yang et al. proposed estimating the overlapping area between the images using phase correction, thereby narrowing the search space for feature detection[32]. This approach is faster than directly stitching using SURF and improves the accuracy of feature matching. However, it still requires images to have enough features since it does not increase the number of detected features as illustrated by the microscopy images with 50% overlap used to present their method. Mohammodi et al. employed SURF as the feature detector and utilized a weighted graph tree to replace camera pose estimation, reducing computational costs[27]. However, to maintain accuracy, they used microscopy images with a 30% overlap. Xu et. al. proposes the enhancement of the features via image edge detection prior to applying SIFT algorithm, to maximize the matching features[33]. However, the method does not address the challenges in microscopy images, when the matching features are sparse and unevenly distributed. Recently, Gan et.al. proposes an optimized SIFT algorithm to address the issue of uneven feature point distribution in the image stitching process, however, the validation is limited to environmental pictures where significant features are available[34]. Preibisch et al. applied a Fourier-Shift Theorem in image stitching with translational variations, however this method does not account for the rotational variations that possibly occur during microscopy image acquisition[35]. MIST is another microscopy image stitching tool[14], which aims for fast stitching but is limited to handle only translational variations due to relying on a similar Fourier-based approach. Moreover, MIST does not apply image blending and the stitched image usually introduces additional noises such as generating seam lines, which can affect the downstream image analysis. Recently developed neural network-based feature detection and matching algorithms, such as SuperGlue[36] and LoFTR[37], point in a new direction for image stitching. However, the current stitching tools based on these algorithms are limited to pairwise stitching and cannot process large volumes of images, which is essential in stitching microscopy images. Recent efforts are being made to improve stitching quality of images beyond significant overlapping. Seo et al. pushes the boundary to 1.15% overlap using a brightness-based normalized cross-correlation method[38]; however, this method can only be applied under the strong assumption that there's no rotation or size scale when acquiring the images.

Here, we propose a feature-based bi-channel aided approach to stitch limited overlapped microscopy images. We demonstrate the proposed method by stitching Atomic Force Microscopy (AFM) generated images of bacterial biofilms via first extracting higher number of matched features from different correlated microscopy imaging channels such as amplitude and then utilizing the information to estimate the alignment of the topographical images. Furthermore, we show that the differentiation of the topographical images along the x-axis provides similar feature information to the amplitude channel image, which generalizes our approach when the amplitude images are not available. We compared the performance of our method with other stitching approaches by experimenting with different biofilm datasets with 10% overlapping and thereby demonstrated the effectiveness of our method. By integrating the proposed workflow, our work aims to improve the balance between cost effective automated microscopy scanning and image stitching accuracy, enabling extraction of accurate physical information and promoting discovery.

**Result and Discussion**

In this paper, we have focused on the study of the early stages of bacterial biofilm formations, where the researchers must capture both the overall structural organization of microbial communities and the fine-scale interactions between individual cells, which are critical for understanding biofilm formation, resistance mechanisms, and responses to environmental changes. AFM[39] has been considered for studying surface attachments it operates effectively in both air and liquid environments, allowing imaging under near physiological conditions[40,41]. Moreover, AFM provides high-resolution imaging and can generate detailed three-dimensional images of surface topography for studying cell structures, cell-to-cell interactions, and fine features like cell walls, flagella and pili[42]. AFM has not only proven to be a powerful tool for nanoscale mechanical and topographical characterization, but it also enables the integration of various non-invasive chemical imaging[43,44] and composition mapping[45], internal hydration properties of single bacterial endospores[46] and properties of outer membrane extensions of bacteria[47]. While AFM can provide detailed insights into individual cells and bio-structures at the nano- to microscale and capture intricate details about the topography and mechanical properties, traditional AFM techniques are still limited by their slow measurement process and small imaging areas (typically<100 μm) due to being restricted by piezoelectric actuator constraints, thereby limiting the use of AFM to study large, millimeter-scale biofilm structures. This scale restriction makes it challenging to gain a representative understanding of biofilm assembly across larger areas. Recently, a large-area AFM method has been developed by Millan-Solsona et al. to enable imaging of millimeter-sized biofilm structures[48]. In order to study the dynamic properties of the biofilm at different stages of formation, the collected large area AFM topographical images need to be stitched together to reconstruct the full image.

AFM techniques provide multi-channel information, offering various data streams that can aid in the image stitching process. In addition, the limited scan area of the device often necessitates capturing an array of overlapping images. These overlapping regions not only enable accurate stitching but also help mitigate common AFM artifacts. Among the most significant artifacts are sample tilt, rotation induced by the piezo drive, hysteresis, thermal drift, and the limited precision of the motorized stage. To ensure a reliable reconstruction, the stitching system must effectively compensate for these distortions while accounting for the degrees of freedom involved. Specifically, in AFM, image stitching must address sample displacement, rotation, and tilt to achieve accurate alignment and seamless integration.

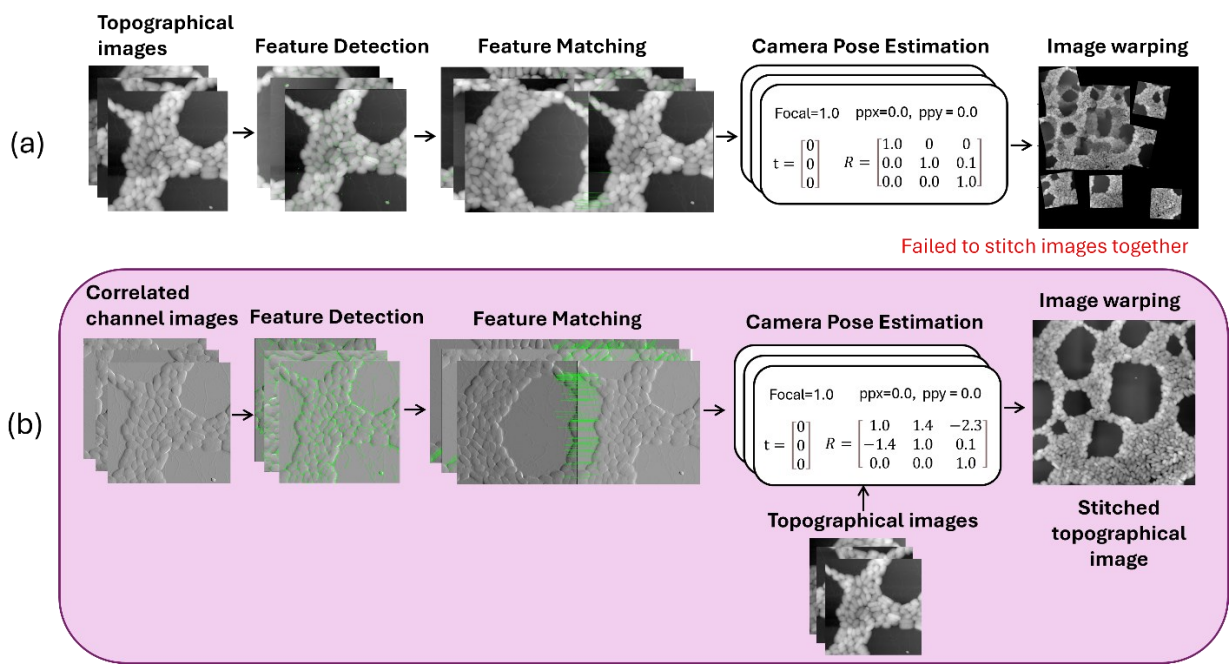

**Figure 1.** Proposed approach of feature-based bi-channel aided image stitching of microscopy images. Here fig. (a) shows the traditional direct stitching of stack of overlapped microscopy topographical images where (b) shows the stitching of the same via our proposed approach. Here in our method, we extract higher matched features (denoted by green lines) from correlated second channel microscopy images (eg. stack of overlapped amplitude or differential images) than extracting matched features directly from the first channel topographical images as in (a). Then, we utilize the insights from a higher number of matched features to improve the camera pose estimation and thereby improve the stitching quality of the topographical images.

In this paper we propose a bi-channel aided microscopy image stitching method based on the underlying feature-based technique. By controlling the input in the feature-based image stitching pipeline, our goal is to obtain a more precise estimate of the rotation, displacement, and tilt variations of the sampled microscopy images by maximizing the extraction of matched features from the limited overlapping areas. **Fig. 1** shows the proposed workflow of bi-channel aided stitching method. In general, as shown in Fig. 1, feature-based stitching involves four key steps: feature detection, feature matching, camera pose estimation, and image warping and blending. In feature detection, distinct patterns or key points from all the scanned microscopy images are identified. Then, the feature matching algorithm extracts the corresponding key points from overlapping areas of adjacent images. In camera pose estimation, we estimate the intrinsic (i.e., focal length and principal point settings) and extrinsic parameters (i.e., rotation and translation) of the camera pose relative to a reference frame. Finally, the warping and blending algorithm applies the estimated image transformations to images and stitches them together seamlessly.

Following our proposed approach of bi-channel aided stitching (Fig. 1b), in this case study of bacterial surface attachment and biofilm formation, topographical images generated from large area-AFM are the sample of interest (first channel) as it captures the morphological details of the attached cells. However, instead of detecting features from topographical images (first imaging channel), we explore other AFM imaging channels that are correlated to the first imaging channel, revealing finer surface details and containing a higher number of detectable features. In this case, this exploration and analysis for identifying the second imaging channel is conducted via domain expert knowledge. Here, to choose the best second imaging channel for feature detection and matching, we find that amplitude images yield more detected and matched features, which enhance the performance in stitching compared to topographical image. To generalize our proposed approach, we aim to investigate other correlated imaging channels when the amplitude images are not available. To eliminate the need for the experimental and storage burden of amplitude images, we investigated further and found that computing the x-axis derivative of the topographical image (first channel) also yields similar feature-rich images like the amplitude images (second channel). Thus, in camera pose estimation, using the highly matched features of correlated microscopy second channel images (eg. amplitude images), we estimate the parameters of the camera pose relative to a reference frame. Since the AFM images are not captured by any conventional camera, the camera pose parameters are treated as an assumed variable used to calculate image transformations. In this step, the algorithm first assigns initial values to the camera pose parameters, then optimizes those using the matched features. Here, these parameters are optimized and estimated from the higher feature-matched second channel such as amplitude channel, which are then used to determine the transformations of the first channel topographical images within a unified coordinate system. In image warping, we align the topographical images together by applying the derived transformation matrix to these images on a per-pixel basis. That is, we first perform feature detection, matching, and camera pose estimation using the second channel images (high featured), and then we apply the estimated transformations to warp the first channel images (sparse featured) together. Finally, the blending algorithm takes the warped topographical images and blends them together seamlessly. It should be noted that our method does not depend on the availability of only amplitude channel images. Rather, we see that the differentiation of the topographical images along the x-axis provides similar feature information to the amplitude channel image, which can be considered as the second channel when the amplitude image is not available. Considering the objectives of having high correlation between first and second channel images and higher detected/matched features of the second channel images, if no better solution is found, then the optimal second channel images is the first channel images.

For all the demonstrated analysis of the proposed approach, we considered the package "Stitching", coded in Python[25]. Stitching provides a Jupyter notebook tutorial that allows us to use each stitching module flexibly. We integrate our idea with the stitching pipeline to perform the AFM topographical images stitching task. We use the amplitude image as input for feature detection. There are four feature detectors available from the Stitching package: ORB, BRISK, AKAZE and SIFT. After exhaustive tuning with different feature algorithms, we chose SIFT as the feature detector with *contrast_Threshold*=0.015, *edge_Threshold* = 15. After extracting key features from all the amplitude images, these features are matched between image pairs using feature matching. There are two feature matching modes available: *homography* and *affine*. *Affine* transformations are computationally cheap and consider scaling, rotation and translation variations whereas *Homography* mode handles one more type of transformation named perspective distortion. However, as these transformations are unlikely to happen in microscopy images as per domain expert knowledge, we use the affine mode for better feature matching. To demonstrate the performance of the proposed method, we considered two datasets of large-area AFM scanned images of early biofilm formation, with each dataset being scanned at 10% overlap.

**Discussion on experimental datasets**

AFM images were acquired using the *DriveAFM* system from *Nanosurf*, which features a motorized platform capable of scanning large areas. The datasets were collected using the methodology described in[48] through a custom Python graphical interface designed to capture image arrays. In this study, we utilized 3 × 3 and 7 × 7 overlapping image arrays, each image of 15 x 15 μm$^2$ and 87 x 87 μm$^2$ respectively with a 10% overlap. For the remainder of the paper, we correspond these to datasets 1 and 2, respectively. The stack of the images in datasets 1 and 2 is provided in Supplementary Materials (**Fig. S1**). All images were acquired in tapping mode, capturing topography, amplitude, deflection, and phase channels. AFM topographic images often contain noise, such as artificial horizontal lines caused by the feedback system, adhesion artifacts, or changes in resonance frequency, which result in abrupt variations between consecutive scan lines. Additionally, AFM measurements require post-processing to correct potential sample tilts, thermal drift, and other distortions. These artifacts can be easily removed using commonly applied techniques, such as mean difference adjustment, median adjustment, surface fitting methods, and flattening techniques. In this study, we used custom functions available in[49] to integrate them into the workflow. The samples used in this study to a time course of surface attachment by ***Pantoea sp. YR343***, a Gram-negative bacterium isolated from the rhizosphere of poplar trees, known for its plant growth-promoting properties[50]. *Pantoea* sp. YR343 is a motile, rod-shaped bacterium with pili and peritrichous flagella, facilitating its interactions within its environment. The strain was grown in minimal medium, as described in[48], on a silicon oxide substrate between 4 and 6 hours for dataset 1 and for 30 minutes for dataset 2. At the time of collection, the coverslip was carefully rinsed in Nanopure water to remove unattached cells, without being allowed to dry at any point. The coverslips were then fixed in 4% paraformaldehyde for 15 minutes and washed once more. Finally, the samples were dried before imaging using pressurized nitrogen.

Before implementation, we also analyze the features of the topography and other imaging channels for these datasets. **Table 1** shows the comparison of the detected and matched features among the topography, amplitude and the differential channels for both datasets. **Figures 2** and **3** show the detected features among the topography and the other imaging channels for a sample image on both datasets respectively. From the figures, it is clear that the images from the amplitude and differential channels contain more features as denoted by the density of green dots.

|  | Average Number of Detected features | | | Average Number of matched features | | |
| --- | --- | --- | --- | --- | --- | --- |
|  | Amplitude channel | Differential channel | Topography channel | Amplitude channel | Differential channel | Topography channel |
| dataset1 | **2746** | **1851** | **375** | **169 (12 matches)** | **149 (12 matches)** | **13 (6 matches)** |
| dataset2 | **2216** | **1455** | **1109** | **168 (56 matches)** | **163 (58 matches)** | **86 (48 matches)** |

**Table 1.** Comparison of the detected and matched features between the amplitude images and topographical images of the *Pantoea* sp. YR343 datasets. The Average Number of matched features column provides the number of total matched image pairs and the average number of matched features for each pair.

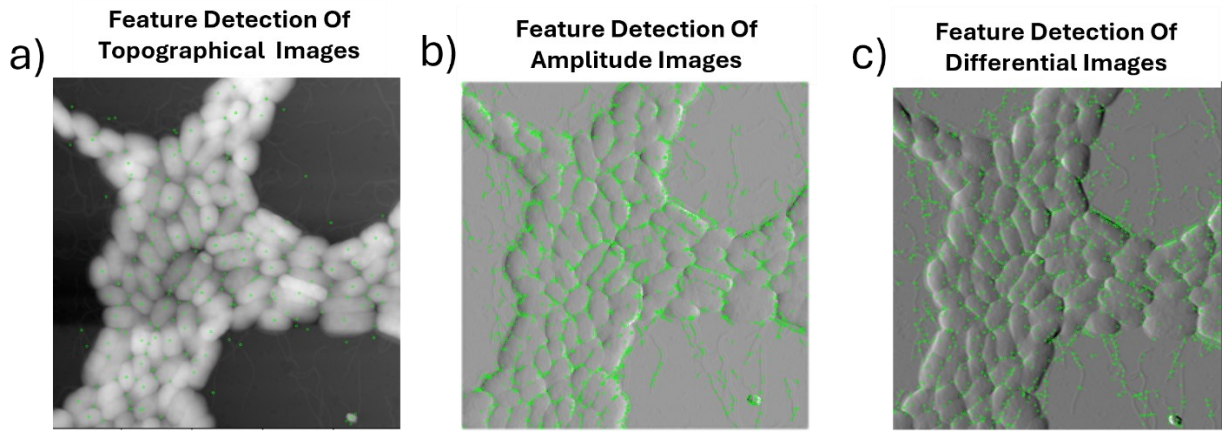

**Figure 2.** Comparison of the detected features between the (a) topographical, (b) amplitude and (c) differential images of the *Pantoea* sp. YR343 dataset 1. Here, we present one sampled scan image from the dataset. The green dots in each of the images are the detected features.

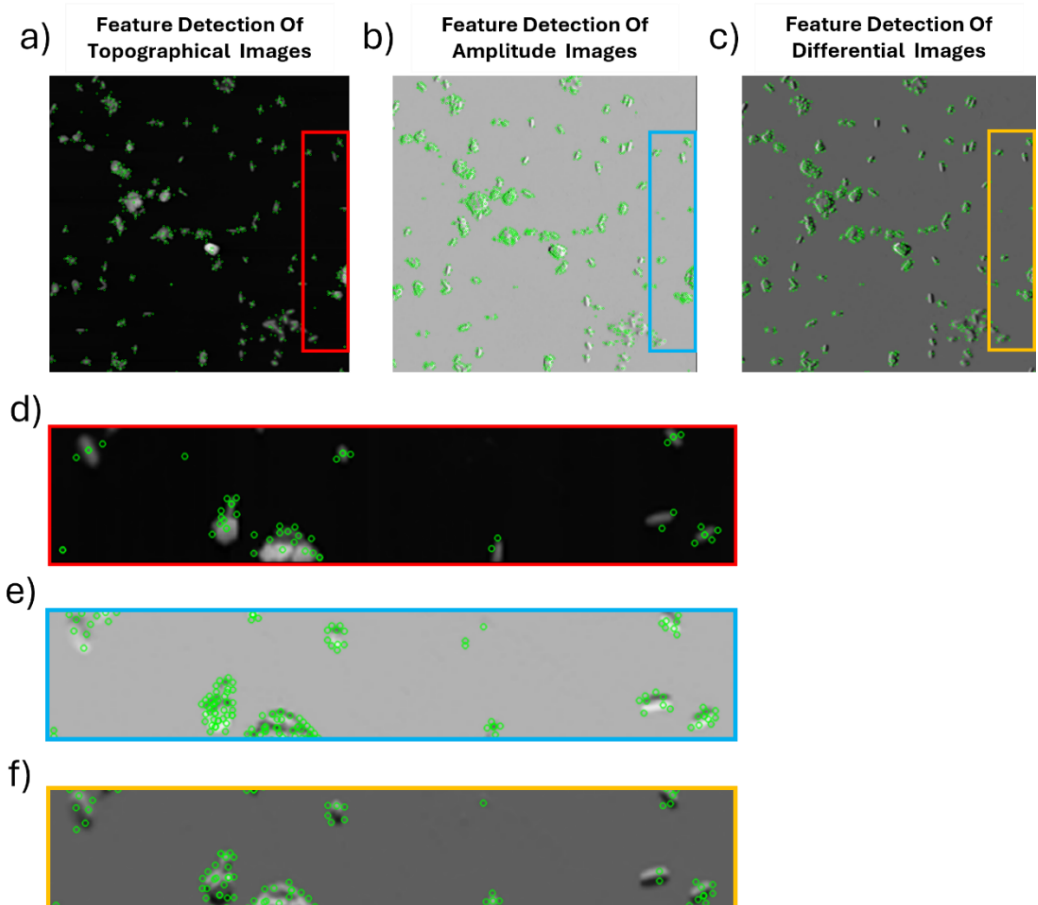

**Figure 3.** Comparison of the detected features between the (a) topographical, (b) amplitude and (c) differential images of the *Pantoea* sp. YR343 dataset 2. Here, we present one sampled scan image from the dataset. Figures (d), (e) and (f) are the zoomed images at the box section provided in (a), (b) and (c) respectively. The green dots in each of the images are the detected features.

**Case Study-Implementation on Dataset 1 of *Pantoea* sp. YR343 sample**

We first implemented our proposed approach of bi-channel aided image stitching to dataset 1 composed of images of early biofilm assembly by *Pantoea* sp. YR343, where the strain was grown on a silicon oxide substrate between 4 and 6 hours. The dataset includes 9 scanned topographical and amplitude images, each image of 15 x 15 μm$^2$ and acquired with a 10% overlap. From table 1, we can see each topographical image yielded an average of 375 detected features, resulting in 6 matched image pairs with approximately 13 matched features per pair. In contrast, the corresponding amplitude images capture finer surface details, with each image averaging 2,736 detected features, this resulted in 12 matched pairs, with each pair averaging 169 matched features. Note that detecting a larger number of features not only improves image matching accuracy but also increases the number of matched pairs, leading to a more complete stitched image. In contrast, the differential images average 1,851 detected features, which is lower than the number of detected features in amplitude images, but significantly higher than the detected features in the topographical images. Even though the number of detected features in the differential images is lower than in the amplitude images, the average number of matched features in the differential images is 149, similar to the 169 matched features in the amplitude images

It should be noted that the stitching performance depends heavily on the matched features, and therefore, the differential images also provide significant key details like the amplitude images. **Fig. 4** provides a visual example of feature matching from the amplitude and topographical images (as denoted by green lines), where amplitude images show a significant larger number of matched features.

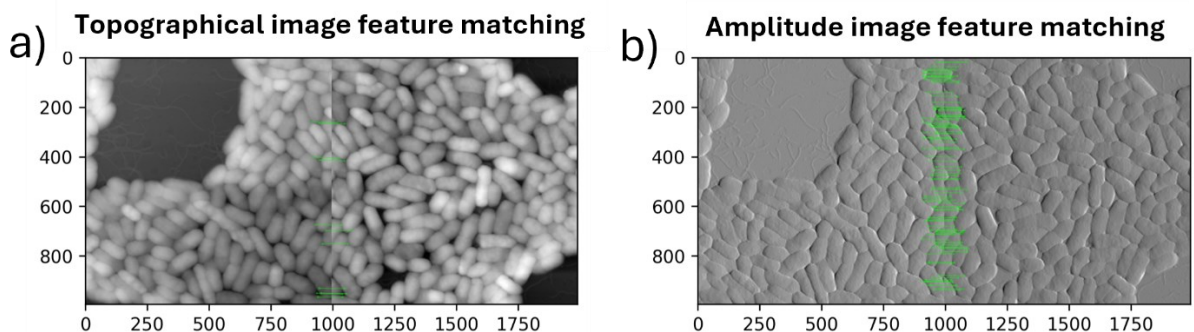

**Figure 4.** Feature matching of (a) topographical and (b) amplitude images of *Pantoea* sp. YR343 dataset 1.

Using the proposed bi-channel aided image stitching method (see Fig. 1), we estimated a transformation matrix from the amplitude images and applied it to the topographical images, resulting in a stitched topographical image. **Figure 5** shows the stitching performance of the proposed method and the comparison with traditional approaches using existing algorithms such as *Stitching*, *Stitch2D* and *MIST*. Here, *Stitching* fails completely to estimate the camera pose and thus it could not provide any solution. From Fig. 5a, we can see *Stitch2D* can only partially stitch images with a 20% overlap where the blank rectangle region is the failed region in the image canvas where the method could not align any images due to potentially sparse matching features of the topographical images (see Table 1). Furthermore, *Stitch2D* fails to stitch any topographical images when the overlap decreases to 10%. *MIST*, on the other hand from Fig. 5b, performed better and could stitch topographical images with a 10% overlap when the overlap percentage is manually specified. However, the stitched topographical image introduced noises such as several seam lines (denoted by red and yellow solid boxes in Fig. 5d and 5e). These seam lines can interfere with subsequent post image analyses, such as image segmentation and object detection, . to study the physical properties of cells during biofilm assembly. Thus, even though *MIST* could be able to stitch all the images, the stitched image is physically undesired as per the domain experts. From Fig. 5c, we can clearly see our method outperformed the traditional approach where it can stitch all the topographical images with 10% overlap without generating any significant noise from stitching (denoted by red and yellow dotted boxes in Fig. 5d and 5e). **Figure 6** shows the comparison of the stitched topographical image via extracting features from amplitude and differential images which are similar to the result using the amplitude channel images.

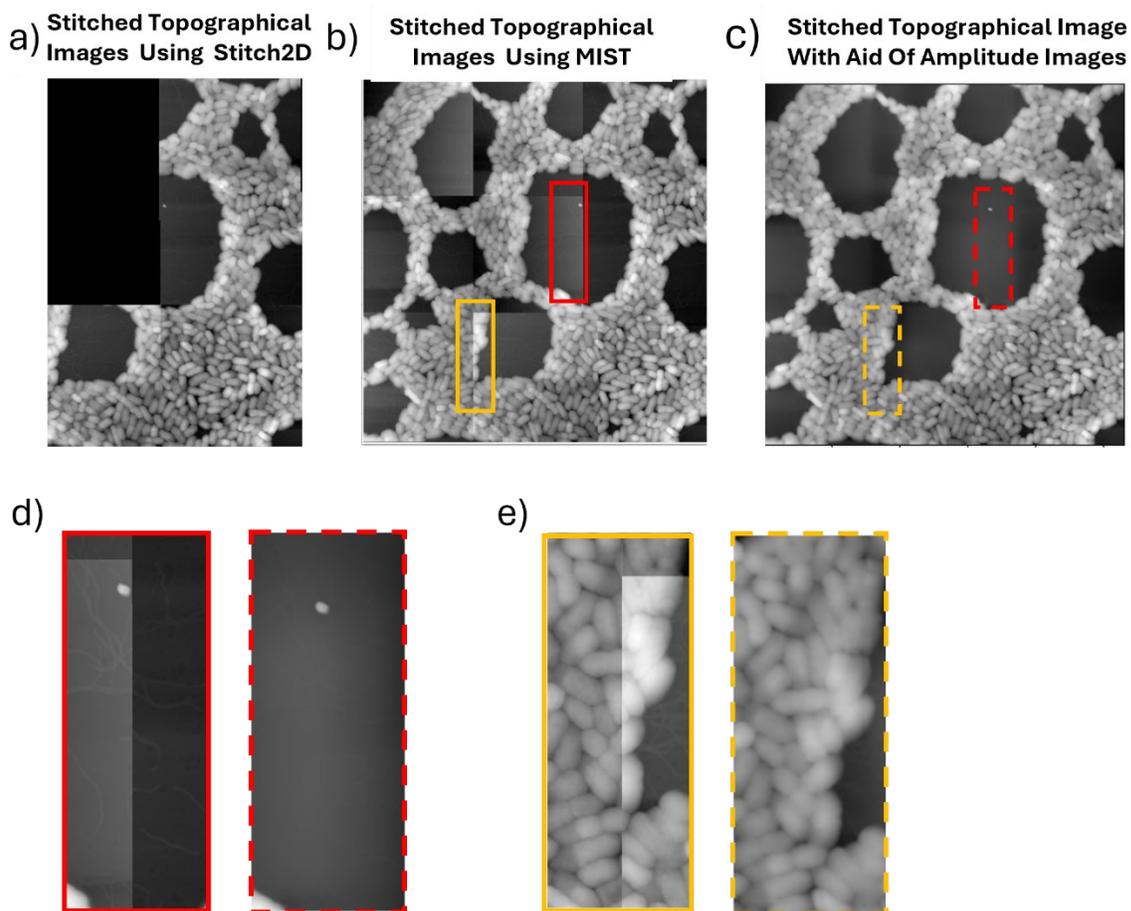

**Figure 5.** Comparison of the stitching of topographical images from *Pantoea* sp. YR343 dataset 1 stitching result using a) *Stitch 2D* with 20% overlap, b) *MIST* with 10% overlap and c) the proposed method with 10% overlap with the aid of amplitude channel images.

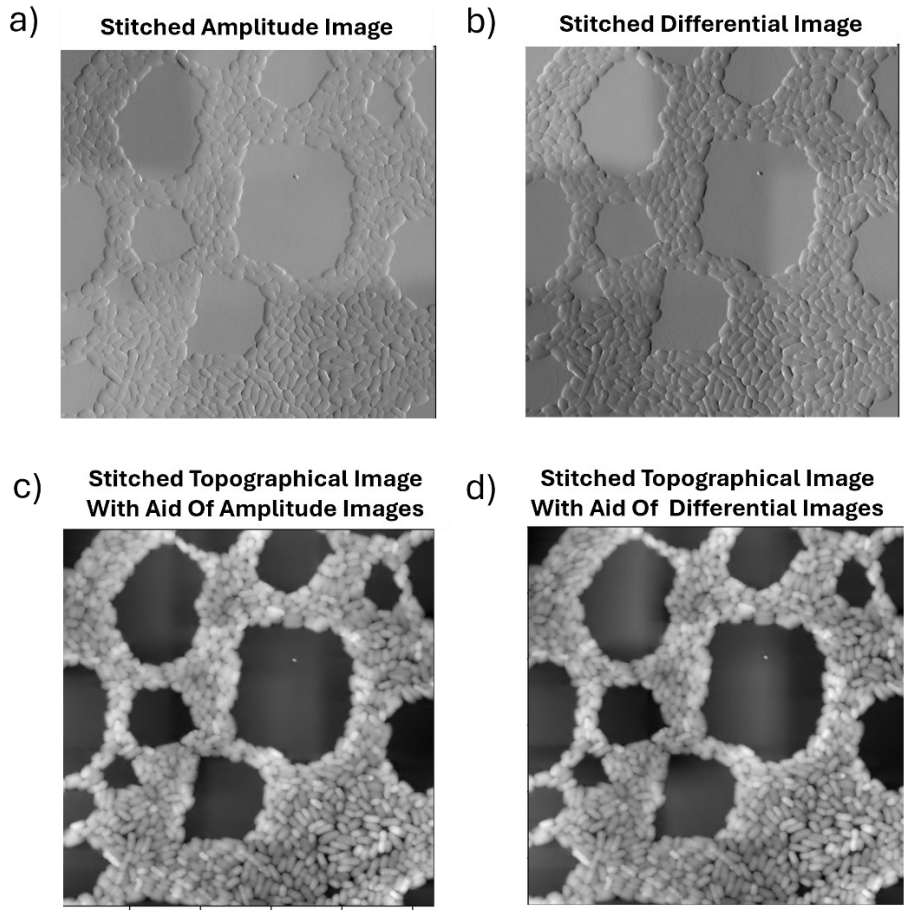

**Figure 6.** Stitching of *Pantoea* sp. YR343 dataset 1 images using different second imaging channel. Here, a) is the stitched amplitude channel image, b) is the stitched differentiation image, c) is the stitched topographical image with aid of the amplitude image, d) is the stitched topographical image with aid of the differentiation of topographical image along x axis.

**Case Study- Implementation on Dataset 2 of *Pantoea* sp. YR343 sample**

Next, we implemented our proposed approach to the biofilm dataset 2 of *Pantoea* sp. YR343, where the strain was grown on a silicon oxide substrate for 30 mins. The dataset includes 49 scanned topographical and amplitude images, each image of 87 x 87 $\mu m^2$ and acquired with a 10% overlap. From table 1, we can see each topographical image yielded an average of 1109 detected features, resulting in 48 matched image pairs with approximately 86 matched features per pair. In contrast, the corresponding amplitude images capture averaging 2,216 detected features and producing 56 matched pairs, each averaging 168 matched features. Here, we see the resulting number of average match features of the differential images is 163, which is close to the same for amplitude images. **Figure 7** shows the stitching performance of the proposed method and the comparison with *Stitching*, *Stitch2D* and *MIST*. Here, *Stitch2D* could not produce any solution as it fails to align the images together. From Fig. 7a, we can see *Stitching* can only partially stitch images with a 10% overlap despite containing a seemingly sufficient number of matched features of the topographical images (see Table 1). This is due to the uneven distribution of matched features and the presence of feature-sparse images, which prevent the algorithm from directly stitching the topographical images. As a result, 15 out of 49 images are missing from the stitched image. *MIST*, on the other hand, performed better and could stitch all the topographical images with 10% overlap when the overlap percentage was manually specified as shown in Figure 7b. However, like in previous analysis, the stitched topographical image introduced more noises such as misalignment (denoted by red solid box in Fig. 7d) and seam lines (denoted by yellow solid box in Fig. 7e), which degrades the stitched image to a unsuitable result. Again from Fig. 7c, we can clearly see our method outperformed the traditional approach where it can stitch all the topographical images with 10% overlap, without any sign of significant misalignment or generation of seam lines. To further investigate the regions where *MIST* fails, we can see our method provides correct alignment and smooth surfaces (denoted by dotted red and yellow boxes in Figs. 7d and 7e respectively) of the stitched topographical image.

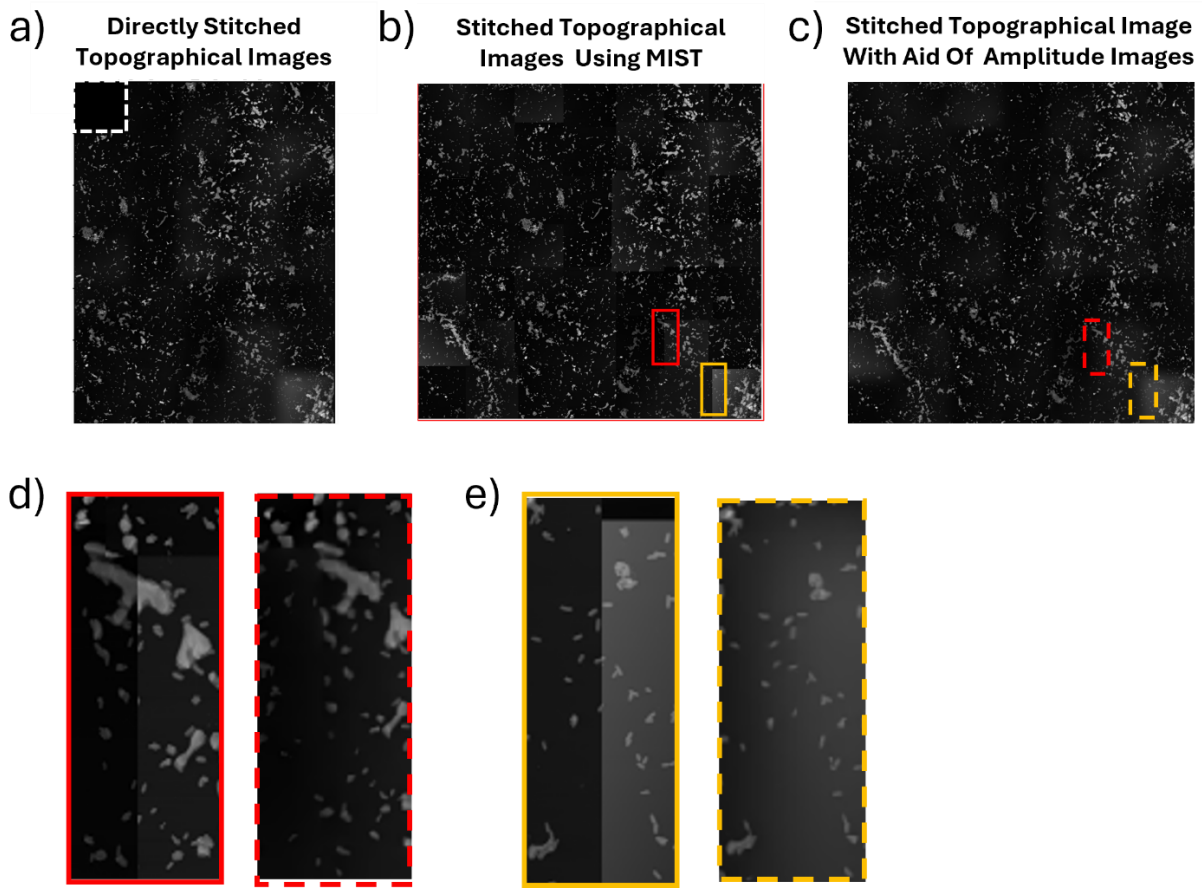

**Figure 7.** Comparison of the stitching of topographical images from *Pantoea* sp. YR343 dataset 2 using a) *Stitching* with 10% overlap, b) *MIST* with 10% overlap and c) the proposed method with 10% overlap with the aid of the amplitude channel images.

Finally, **Figure 8** shows the comparison of the stitched topographical image via extracting features from amplitude and differential imaging channels. Similar to previous analysis, both stitched images are very similar, suggesting that our method can be generalized to other image types, serving as a valuable augmentation strategy for microscopy image stitching tasks.

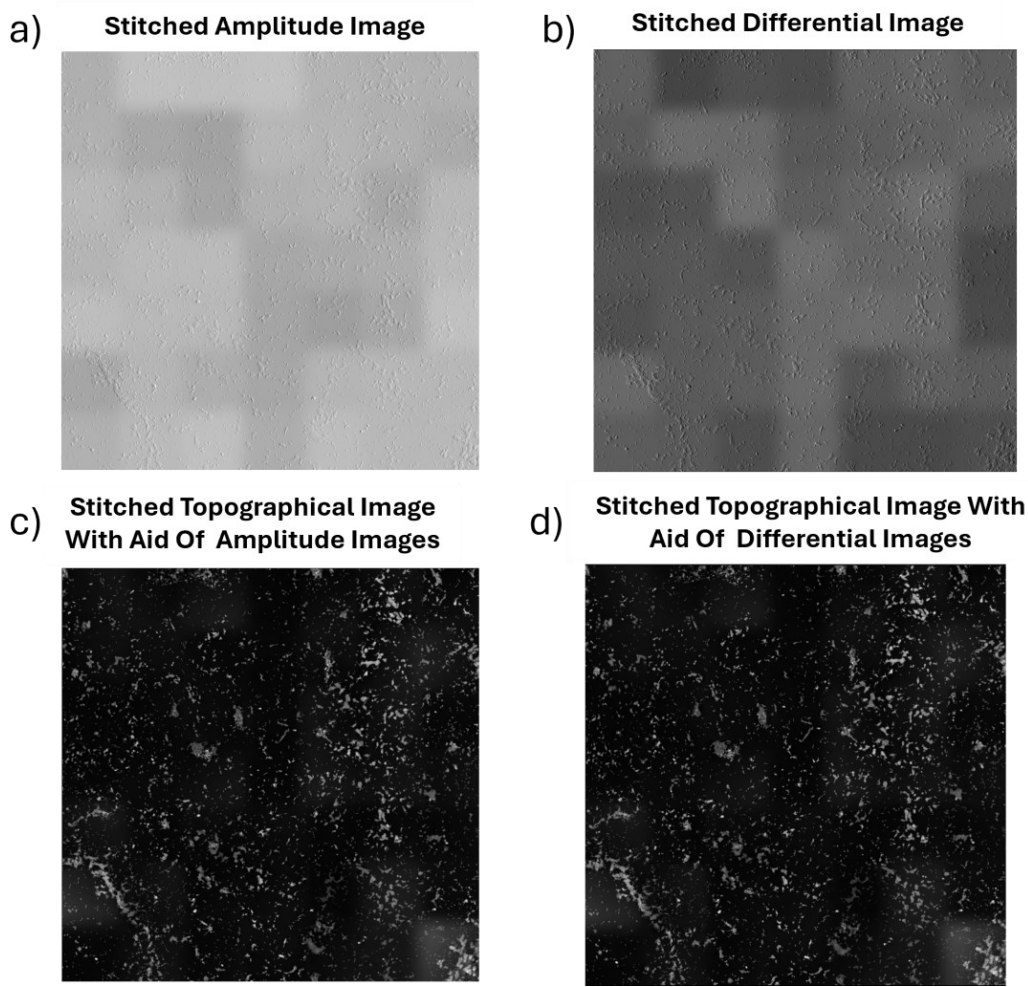

**Figure 8.** Stitching of *Pantoea* sp. YR343 dataset 2 images using different second imaging channel. Here, a) is the stitched amplitude channel image, b) is the stitched differentiation image, c) is the stitched topographical image with aid of amplitude image, d) is the stitched topographical image with aid of the differentiation of topographical image along x axis.

**Summary**

In summary, we propose a microscopy imaging bi-channel aided stitching method to aim to gather larger volume of information needed to stitch multiple images via maximizing extraction of matching features. Here, the first imaging channel is the stack of images of interest to stitch together and the second channel is the stack of images, highly correlated to the first channel, where higher volume of detected and matched features between adjacent image pairs can be extracted. In other words, our proposed pipeline is designed to stitch any microscopy image of interest as

follows- 1) performing feature detection and matching from the rich-featured correlated image channels (such as amplitude or differential images), 2) then using the feature information to optimize the camera pose estimation, and 3) then we apply the estimated transformations to warp the sparse-featured image of interest to stitch them together. We see that the existing stitching tools generally work well when the image of interest is either less complex such as environmental images or with significant overlapping (eg. 30%) of image pairs. However, real microscopy experimental data is more complex with undistributed and sparse features and additional artifact noise. Furthermore, data acquisition with such an overlapping percentage can be extremely challenging in terms of experimental cost. We demonstrated our method to stitch two sets of complex biofilm assembly images with limited (e.g. 10%) overlaps where our method outperformed significantly than the traditional stitching tools like *Stitching*, *Stitch2D* and *MIST*. We believe that our method can be generalized to stitch any microscopy images (such as STM, STEM etc.) to study large-scale material structures, and benefit the experimentalist to avoid erroneous analysis and discovery due to incorrect stitching. In the future, we aim to automate the optimal selection of the second imaging channel by designing a hybrid objective metric of quantifying mean correlation between stacks of images from first and second imaging channel, and quantifying the mean detected and matched features (as per Table 1) of the candidate for second imaging channel. We aim to also integrate the method with this hybrid metric into microscopy for real-time automated stitching. Future work also seeks to extend the method towards 1) adapting deep-learning framework for improved detection of matched features and 2) adding workflow to gather prior knowledge of the features from pretraining with existing microscopy image data from correlated material systems.

**Additional Information:**
See the supplementary material for additional analysis and figures related to the research.

**Acknowledgements:**
This work (A.B and H.Z) was supported by the University of Tennessee startup funding. The authors acknowledge the use of facilities and instrumentation at the UT Knoxville Institute for Advanced Materials and Manufacturing (IAMM) and the Shull Wollan Center (SWC) supported in part by the National Science Foundation Materials Research Science and Engineering Center program through the UT Knoxville Center for Advanced Materials and Manufacturing (DMR-2309083). Work by R.M, S.R.B., M.C, L.C and J.L.M was supported by the U.S. Department of Energy, Office of Science FWP ERKCZ64, Structure Guided Design of Materials to Optimize the Abiotic-Biotic Material Interface, as part of the Bio preparedness Research Virtual Environment (BRaVE) initiative. AFM measurements and sample preparation were conducted as part of a user project at the Center for Nanophase Materials Sciences (CNMS), which is a US Department of Energy, Office of Science User Facility at Oak Ridge National Laboratory. The authors would like to sincerely thank Scott Retterer for his generous assistance in the acquisition of funding for this research.

**Conflict of Interest:**
The author confirms there is no conflict of interest.
**Code and Data Availability Statement:**
The analysis reported here along with the code is summarized in Notebook for the purpose of tutorial and application to other data and can be found in <https://github.com/arpanbiswas52/Stitching_AFMimage>

Supplementary Materials of the paper titled
# A Bi-channel Aided Stitching of Atomic Force Microscopy Images

Huanhuan Zhao[1], Ruben Millan-Solsona[2], Marti Checa[2], Spenser R. Brown[3], Jennifer L. Morrell-Falvey[3], Liam Collins[2], Arpan Biswas[4]

[1]Bredesen Center for Interdisciplinary Research, University of Tennessee, Knoxville, USA, 37996
[2]Center for Nanophase Materials Sciences, Oak Ridge National Laboratory, Oak Ridge, USA, 37830
[3]Biosciences Division, Oak Ridge National Laboratory, Oak Ridge, USA, 37830
[4]University of Tennessee-Oak Ridge Innovation Institute, University of Tennessee, Knoxville, USA, 37996

**Appendix A. Datasets used in the paper.**

(a) Dataset 1: 3*3 10% overlap

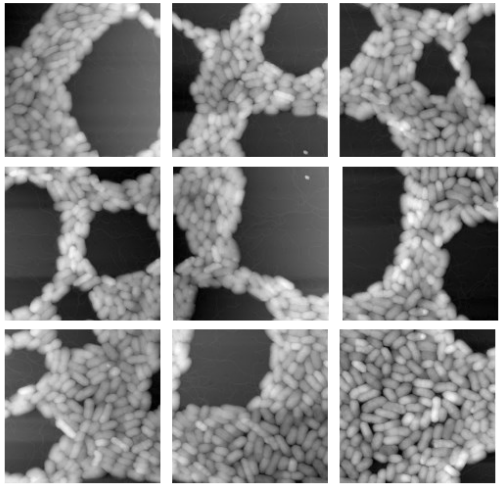

(b) Dataset 2: 7*7 10% overlap

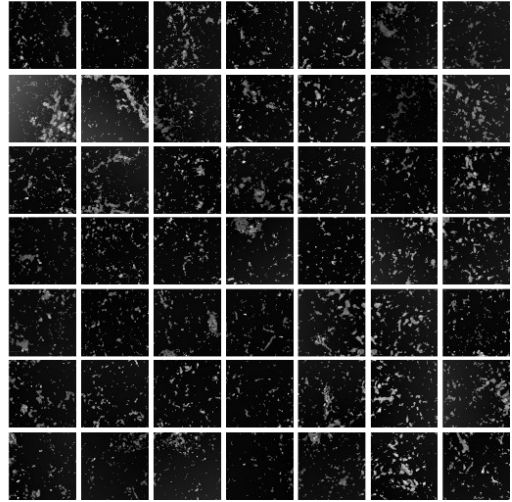

**Figure S1**. (a) is dataset 1 contains 9 images, each of 15 x 15 $\mu m^2$ and (b) is dataset 2 contains 49 images, each of 87 x 87 $\mu m^2$, images in (b) are not arranged in a stitching sequence. This highlights that arranging such microscopy images manually, as in dataset 2, is too complex and time-consuming, thus not a feasible approach. Notably, the proposed method does not require prior information such as percentage of image overlapping or ordering.

**Appendix B. Implementation on Dataset 3 of *Pantoea* sp. YR343 sample, considering 5x5 image arrays.**

The dataset 3 contains 5x5 image arrays, each of which is 87 x 87 μm$^2$ with 10% overlap. Here, the strain was grown in minimal medium, as described in[48], on a silicon oxide substrate for 60 minutes.

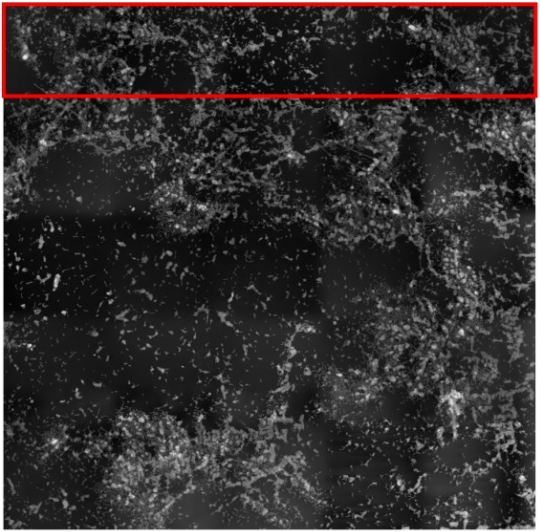
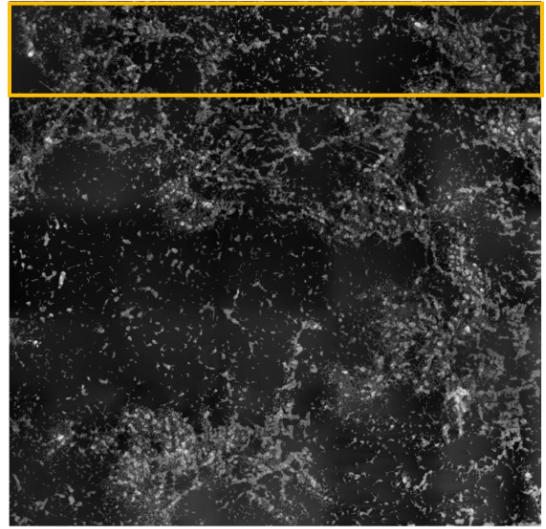
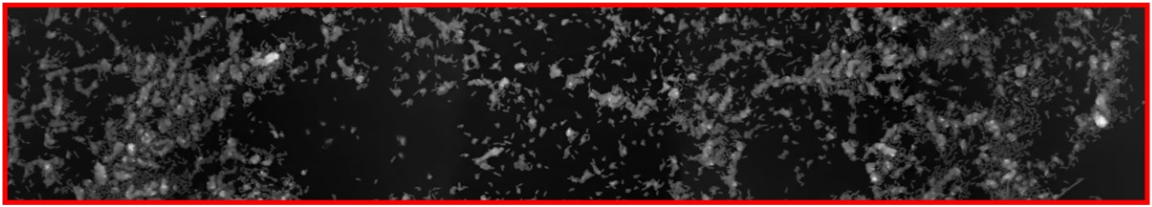
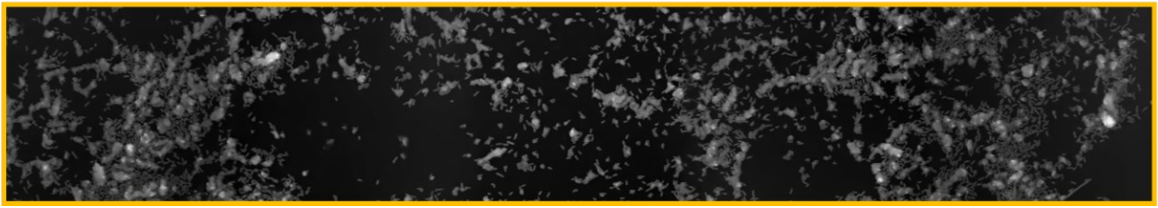

**Figure S2.** Here, we can see an example where the topography images contain rich features, making direct stitching of the images feasible. (a) is the stitched topographical image from directly stitching. (b) is the stitched topographical image from proposed method. (c) is the zoomed in picture from the red rectangular area in (a), (d) is the zoomed in area from the yellow rectangular area in (c). The Structure Similarity Index (SSIM) score of (a) and (b) is 0.99 which indicates these two stitched images are highly similar. We see our proposed method also provides a comparable stitching performance with no sign of performance degradation. Thus, our method either improves stitching or provides at least similar performance, depending on the volume of features in the first channel of images.